%% file: hivetool.tex
\documentclass[submission]{eptcs}
\providecommand{\event}{PDMC 2011}

\newif\ifapp\appfalse
\appfalse
\newif\ifpaper\paperfalse
\papertrue

\input{preamble}

\usepackage{latexsym}
\usepackage{amssymb}
\usepackage{epsfig} 
\usepackage{theorem}
\usepackage{graphics}
\usepackage{amsmath}
\usepackage{url}
\usepackage{color}
\usepackage{dsfont}
\usepackage{multirow}
\usepackage{float}
\usepackage{algorithmic}
\usepackage{wrapfig}
\usepackage{subfigure}
\usepackage{array}
\usepackage{doi}
\usepackage[mathcal]{euscript}

\usepackage[T1]{fontenc}

\floatstyle{ruled}
\newfloat{algorithm}{tbph}{al}
\floatname{algorithm}{Algorithm}

\newcommand{\LTS}{\mbox{\sc Lts}}

\newcommand{\LTSs}{\mbox{\sc Lts}s}

\newcommand{\C}{\mathfrak{C}}
\newcommand{\Ps}{\mathfrak{P}}
\newcommand{\M}{\mathfrak{M}}
\newcommand{\R}{\mathfrak{R}}

\newcommand{\sint}{s_{\it in}}

\newcommand{\BFS}{\mbox{\sc Bfs}}
\newcommand{\DFS}{\mbox{\sc Dfs}}

\newcommand{\ISS}{\mbox{\sc Iss}}
\newcommand{\ISSs}{\mbox{\sc Iss}s}

\newcommand{\LTSmin}{\mbox{\sc LTSmin}}

\newcommand{\HIVE}{\mbox{\sc Hive}}

\newcommand{\tabsize}{\scriptsize}
\newcommand{\tabsizetwo}{\tiny}

\ifpaper
\title{The \textsc{Hive} Tool for Informed Swarm State Space Exploration}
\author{Anton Wijs\thanks{Supported by the Netherlands Organisation for Scientific Research (NWO) project 612.063.816 {\it Efficient Multi-Core Model Checking.}}
\institute{Eindhoven University of Technology, 5612 AZ Eindhoven, The Netherlands}
\email{A.J.Wijs@tue.nl}
}
\def\titlerunning{The \textsc{Hive} Tool for Informed Swarm State Space Exploration}
\def\authorrunning{A.J. Wijs}
\begin{document}
\maketitle

\begin{abstract}
Swarm verification and parallel randomised depth-first search are very effective parallel techniques to hunt bugs in large state spaces. In case bugs are absent, however, scalability of the
parallelisation is completely lost. In recent work, we proposed a mechanism to inform the workers which parts of the state space to explore. This mechanism
is compatible with any action-based formalism, where a state space can be represented by a labelled transition system. With this extension, each worker
can be strictly bounded to explore only a small fraction of the state space at a time. In this paper, we present the \HIVE\ tool together with two search algorithms
which were added to the \LTSmin\ tool suite to both perform a preprocessing step, and execute a bounded worker search. The new tool is used to coordinate informed swarm explorations, and the two new \LTSmin\ algorithms are employed
for preprocessing a model and performing the individual searches.
\end{abstract}

\section{Introduction}
\label{sec:intro}

In explicit-state model checking (MC), it is checked whether a given system specification yields a given temporal property. This is done by exploring the so-called {\it state space} of the specification, which is a directed graph describing explicitly all potential behaviour of the system. Since state space exploration algorithms often need to keep track of all explored states in order to efficiently perform the MC task,\footnote{A Depth-First Search can in principle be performed by just using a stack, but this means that the MC task can often not be performed in linear time (depending on the structure of the state space).} and since state spaces can be very large, for many years, the amount of available memory in a computer has been the most important bottleneck for MC.

In recent years, however, the increase of available memory in state-of-the-art computers has continued to follow Moore's Law~\cite{moore}, while the increase of their processors' speed no longer has. For MC, this means that large state spaces can be stored in memory, but the time needed to explore them is impractically long, hence a {\it time explosion problem} has emerged. This can be mitigated by developing {\it distributed} exploration algorithms, in which a number of computers in a cluster or grid are used to perform an exploration. Many of those algorithms use a partitioning function to assign states to workers, and require frequent synchronisation between these workers, see e.g.~\cite{divine10,preach,ltsmintool,Garavel2006,lerda.sista.distributedspin}.

{\it Swarm verification}~\cite{holzmann.swarm2} (SV) (and {\it parallel randomised Depth-First Search}~\cite{dwyer.elbaum.parallelrandomized}) are recent techniques to perform state space exploration in a so-called {\it embarrassingly parallel}~\cite{foster.parallel} way, where the individual workers never need to synchronise with each other. In SV, each worker starts at the initial state and performs a search based on Depth-First Search (\DFS). The direction of a worker is determined by a given successor ordering strategy. As the direction of a \DFS\ depends on the fact that a stack is used to order successor states (i.e.\ a Last-In-First-Out strategy), changing this ordering directly influences the direction of the search. By providing each worker a unique strategy, they will explore different parts of the state space first. With this method, some states may be explored multiple times by different workers, but if the property does not hold, any bug states present are likely to be detected very quickly, due to the diversity of the searches, which often means that the whole state space does not have to be explored.

However, if a property holds, each worker will exhaustively explore the whole reachable state space, which means that the benefits of parallelisation are completely lost. Recently, we proposed a mechanism to bound each worker to a particular reachable strict subset of the set of reachable states, in such a way that together, the workers explore the whole state space~\cite{wijs.isv}. This mechanism is compatible with any action-based formalism such as $\mu$CRL~\cite{gp95}, where each transition in a state space is labelled with some action name corresponding with system behaviour. In this paper, we explain how the {\it Heuristics Instructor for parallel VErification} (\HIVE) tool, which resulted from~\cite{wijs.isv}, works in practice.
Section~\ref{sec:ISV} presents the functionality of the \HIVE\ tool together with some new algorithms implemented in the \LTSmin\ tool suite~\cite{ltsmintool}. How all these have been implemented and how the resulting tools can be used is explained in Section~\ref{sec:impl}. In Section~\ref{sec:exp}, experimental results are discussed. Finally, conclusions and pointers for possible future work are given in Section~\ref{sec:conc}.

\section{The Informed Swarm Exploration Technique}
\label{sec:ISV}

\begin{table}[t]
\centering
\caption{The four major functionalities of ISV}
\begin{tabular}{| l | l |}
\hline
\tabsize \textbf{\textsc{LTSmin}} & \tabsize \textbf{\textsc{Hive}}\\
\hline
\hline
\tabsize {\bf P1.\ Trace-counting \textbf{\textsc{Dfs}}}: Constructs $\sprocess'$ with ${\it tc}(s) = \min(1, \Sigma_{s'\in\Next'} {\it tc}(s'))$. & \tabsize {\bf F1.\ Trace selection}: select a swarm trace $\sigma$ for worker\\
\tabsize {\bf F2.\ Informed Swarm Search (\textbf{\textsc{Iss}})}: Search of $\sprocess$ restricted to $\sigma$ & \tabsize {\bf F3.\ Update swarm set:} remove inspected traces\\
\hline
\end{tabular}
\label{tab:functions}
\end{table}

\paragraph*{The Setting}
The so-called {\it Informed} SV technique (ISV) implemented in \HIVE\ and \LTSmin\ is applicable if three conditions are met: (1) A system specification $\Ps$ should be an implicit description of a {\it Labelled Transition System} (\LTS) $\sprocess$. An \LTS\ $\sprocess$ is a quadruple $(\states, \actions, \transitions, \sint)$, where $\sint$ is the initial state, $\states$ is the set of states reachable from $\sint$, $\actions$ is a set of transition labels (actions), and $\transitions:\states \times \actions \times \states$ is the set of transitions between states. With $s \step{A} t$, $A \subseteq \actions$, we say that there exists an $\ell \in A$ such that $(s,\ell,t) \in \transitions$. The reflexive transitive closure of $\step{}$ is denoted as $\tstep{}{^*}$. In on-the-fly state space exploration, $\sint$ and $\actions$ are known a priori, but $\states$ and $\transitions$ are not, and a next-state function $\Next:\states \to 2^\states$ provides the set of successors of a given state. A state $t$ is the successor of a state $s$ iff $s\step{\actions} t$. $\Next$ is used to construct $\states$ and $\transitions$, starting at $\sint$. In the following, we use the notation $\Next \mid A$, with $A\subseteq\actions$, to denote $\Next$ restricted to a set of transition labels $A$, i.e.\ $\Next\mid A (s) = \{ s'\in\states \mid \exists \ell \in A . (s,\ell,s')\in\transitions\}$. Clearly, $\Next \mid \actions = \Next$. Finally, a sequence of actions $\langle \ell_0, \ell_1, \ldots \rangle$ describes all transition sequences (traces) through an \LTS\ $\sprocess$ with $\sint \step{\{\ell_0\}} s_0\step{\{\ell_1\}} s_1 \cdots$ for some $s_0,s_1,\ldots$. If $\sprocess$ is label-deterministic, i.e.\  for all $s,t\in\states, \ell\in\actions$ with $s\step{\ell} t$, there does not exist a state $t' \neq t$ with $s\step{\ell} t'$, such an action sequence corresponds to a single trace. Here, we assume that all \LTSs\ are label-deterministic. If this is not the case, relabelling of some transitions can resolve this.

(2) $\Ps$ should consist of a {\it finite number $n>1$ of process descriptions (e.g.\ process algebraic terms) in parallel composition}. This is the case for any concurrent system. (3) At least some of these processes in parallel composition, i.e.\ a subsystem, should yield {\it finite behaviour}, hence only finite traces. This is not a strict requirement, but if it is not met, then the method relies on bounded analysis of the subsystem, and it does not automatically guarantee anymore that all reachable states are visited.

\paragraph*{ISV}
Say that a specification describes a system of concurrent processes $\Ps = \{P_0, \ldots, P_n\}$, with $n\in\natdom$. ISV exploits the fact that parallel composition is a major cause for state space explosion, and that \LTS\ $\sprocess$ of $\Ps$ is the synchronous product of \LTSs\ $\sprocess_i = (\states^i, \actions^i, \transitions^i, \sint^i)$ of the $P_i$ ($0 \leq i \leq n$), restricted by some synchronisation rules between processes, given by a symmetric function $\C$. E.g.\ $\C(\ell,\ell')=\ell''$ states that if actions $\ell$ and $\ell'$ can be performed by different processes, then the result is action $\ell''$ in the system. For the formal details, see~\cite{wijs.isv}. We assume that the $\actions^i$ are disjoint (if this is not the case, then some rewriting can resolve this)\footnote{Strictly speaking, a weaker requirement suffices~\cite{wijs.isv}.} and that no action is involved in more than one rule defined by $\C$ (either as an input, or as a result). All this implies that for any $\ell\in\actions$, it can be determined whether or not it stems from some behaviour of a particular process $P_i$. Say that $\actions_c \subseteq \actions$ is the set of actions stemming from synchronisations, and that $\actions^i_c \subseteq \actions^i$ is the set of actions of $\sprocess_i$ which are forced to synchronise with other actions, then $\actions = (\bigcup_{i \leq n} \actions^i \setminus \actions^i_c) \cup \actions_c$. Now, for any $A\subseteq \actions$, we can define $\M(A)$ as $\{ \ell'' \in \actions_c \mid \exists \ell\in A, \ell' \not\in A. \C(\ell,\ell') = \ell'' \}$, which is the set of actions resulting from synchronisation involving one action in $A$. Finally, the assumptions about $\C$ allow us to define a relabelling function $\R$ as follows: $\R(\{\ell\}) = \{\ell''\}$ iff there exists an $\ell'$ such that $\C(\ell,\ell') = \ell''$, and $\R(\{\ell\}) = \{\ell \}$, otherwise. 


\begin{wrapfigure}[22]{l}{0.66\textwidth}
\subfigure{
\includegraphics[scale=0.245]{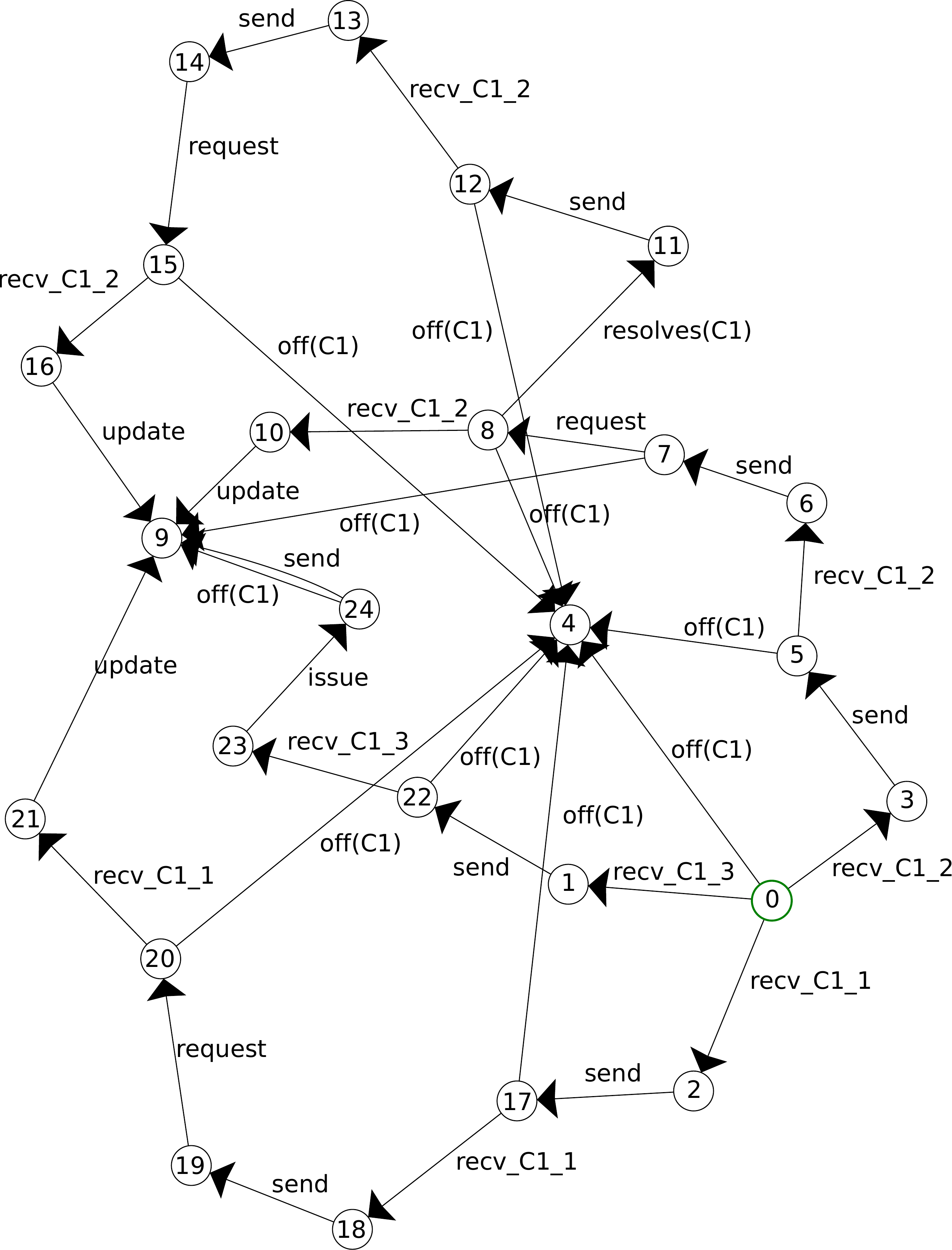}
}
\subtable{
\begin{tabular}[b]{| c | c |}
\hline
\multicolumn{2}{|c|}{{\tabsizetwo \bf Trace counts}}\\
\hline
\tabsizetwo {\it State} & \tabsizetwo {\it tc} \\
\hline
\hline
\tabsizetwo 0 & \tabsizetwo 14 \\
\tabsizetwo 1 & \tabsizetwo 3 \\
\tabsizetwo 2 & \tabsizetwo 3 \\
\tabsizetwo 3 & \tabsizetwo 7 \\
\tabsizetwo 4 & \tabsizetwo 1 \\
\tabsizetwo 5 & \tabsizetwo 7 \\
\tabsizetwo 6 & \tabsizetwo 6 \\
\tabsizetwo 7 & \tabsizetwo 6 \\
\tabsizetwo 8 & \tabsizetwo 5 \\
\tabsizetwo 9 & \tabsizetwo 1 \\
\tabsizetwo 10 & \tabsizetwo 1 \\
\tabsizetwo 11 & \tabsizetwo 3 \\
\tabsizetwo 12 & \tabsizetwo 3 \\
\tabsizetwo 13 & \tabsizetwo 2 \\
\tabsizetwo $\cdots$ & \tabsizetwo $\cdots$ \\
\tabsizetwo 24 & \tabsizetwo 2 \\
\hline
\end{tabular}
}
\subtable{
\begin{tabular}[b]{| c |}
\hline
{\tabsizetwo \bf Trace \#10}\\
\hline
\hline
\tabsizetwo $0 \rightarrow [0,14\rangle$\\
\tabsizetwo $3 \rightarrow [6,13\rangle$\\
\tabsizetwo $5 \rightarrow [6,13\rangle$\\
\tabsizetwo $6 \rightarrow [7,13\rangle$\\
\tabsizetwo $7 \rightarrow [7,13\rangle$\\
\tabsizetwo $8 \rightarrow [7,12\rangle$\\
\tabsizetwo $11 \rightarrow [9,12\rangle$\\
\tabsizetwo $12 \rightarrow [9,12\rangle$\\
\tabsizetwo $13 \rightarrow [10,12\rangle$\\
\tabsizetwo $14 \rightarrow [10,12\rangle$\\
\tabsizetwo $15 \rightarrow [10,12\rangle$\\
\tabsizetwo $4 \rightarrow [10,11\rangle$\\
\hline
\multicolumn{1}{c}{
\vspace{1.92cm}}
\end{tabular}
}
\vspace{-0.6cm}
\caption{The \LTS\ of an iPod process, with weights, and trace nr.\ 10}
\label{fig:ipod}
\end{wrapfigure}
Four basic functionalities are required to perform ISV in practice. These are listed in Table~\ref{tab:functions}; a preprocessing step (P1) involving the analysis of a defined subsystem yielding finite behaviour, and three techniques for the three major phases of ISV (F1-3). P1 and F2 require two new search algorithms, which have been implemented in \LTSmin. F1 and F3 have been implemented in the new stand-alone \HIVE\ tool. The general procedure to perform an ISV is as follows: First, the user selects a strict subsystem $\Ps' \subset \Ps$ which is guaranteed to yield finite behaviour (again, alternatively, this behaviour is bounded in the ISV, but then, the overall search could be non-exhaustive). In P1, the \LTS\ $\sprocess'=(\states',\actions',\transitions',\sint')$, described by $\Ps'$, is constructed and saved to disk, together with a weight function ${\it tc}:\states' \to \natdom$. For this, we have extended the \DFS\ implementation in \LTSmin, as described in Alg.~\ref{alg:tracecounting}. The {\it tc} function (see also Table~\ref{tab:functions}) assigns $1$ to deadlock states, i.e.\ if $\Next' (s) = \emptyset$, and the sum of all the successor weights to any other state (note that $\Next'$ is the next-state function of $\sprocess'$).

This allows efficient reasoning about the traces through $\sprocess'$; the number of traces represented by a trace prefix from $\sint$ to some $s\in\states'$ equals ${\it tc}(s)$.
E.g., Fig.~\ref{fig:ipod} shows a simplified acyclic \LTS\footnote{The actual \LTS\ of this example, in which the actions are extended with some data parameters, consists of $547$ states.} of an iPod process as part of the DRM protocol specification from~\cite{drm}, with part of the definition of {\it tc}. With this, if we sort the states based on their numbering (which was assigned by the \DFS), then each trace can be uniquely referred to with a natural number: note that the number of possible traces is $14$, which corresponds with ${\it tc}(0)$. Say that we want to identify trace 10 (shown in Fig.~\ref{fig:ipod}). State $0$ has successors $\{1,\ldots, 4\}$. Sorted by increasing state number, we first consider state $1$; since ${\it tc}(1)=3$, we conclude that trace $\langle \textsf{recv\_C1\_3} \rangle$ represents traces $0$ to $2$, i.e.\ $3$ traces, starting at trace $0$.

\begin{wrapalgorithm}[13]{r}{0.35\textwidth}
\caption{Trace-counting \DFS}
{\scriptsize
\begin{algorithmic}\label{alg:tracecounting}
\REQUIRE $\Ps' \subset \Ps$, $\sint'$, $\actions'$
\ENSURE $\sprocess'$ and ${\it tc}:\states'\rightarrow\natdom$ are constructed
\STATE $\Closed \leftarrow \emptyset$
\STATE ${\it tc}(\sint) \leftarrow {\it dfs}(\sint)$
\end{algorithmic}
{\bf \textit{dfs(s) =}}
\begin{algorithmic}\label{alg:tracedfs}
\IF {$s \not\in \Closed$}
\STATE ${\it tc}(s) \leftarrow 0$
\FORALL {$s' \in \Next' (s)$}
\STATE ${\it tc}(s) \leftarrow {\it tc}(s) + {\it dfs}(s')$
\ENDFOR
\IF {$\Next' (s) = \emptyset$}
\STATE ${\it tc}(s) \leftarrow 1$
\ENDIF
\STATE $\Closed \leftarrow \Closed \cup \{s\}$
\ENDIF
\RETURN ${\it tc}(s)$
\end{algorithmic}
}
\end{wrapalgorithm}

We also have ${\it tc}(2)=3$, therefore $\langle \textsf{recv\_C1\_1} \rangle$ represents traces $3$ to $5$. Similarly, $\langle \textsf{recv\_C1\_2} \rangle$ represents traces $6$ to $12$, and $\langle \textsf{off(C1)} \rangle$ represents trace $13$. This means that $\langle \textsf{recv\_C1\_2} \rangle$ is a prefix of trace $10$. Since state $3$ only has state $5$ as a successor, clearly $\langle \textsf{recv\_C1\_2}, \textsf{send} \rangle$ is also a prefix (this agrees with ${\it tc}(5)=7$: all $7$ traces represented by $\langle\textsf{recv\_C1\_2} \rangle$ are also represented by $\langle \textsf{recv\_C1\_2}, \textsf{send} \rangle$). In this fashion, the complete trace can be constructed following the states listed on the right of Fig.~\ref{fig:ipod}. This principle is used for trace selection in \HIVE\ (F1). In ISV, each worker is bounded by a trace through $\sprocess'$, given by \HIVE\ (this will be explained next). Therefore, each trace represents a worker job to be performed, and $\sprocess'$ represents the set of jobs. From a trace $\langle \ell_0, \ldots, \ell_n \rangle$ through $\sprocess'$ ($n\in\natdom$), a so-called {\it swarm trace} $\sigma=\langle \R(\ell_0), \ldots, \R(\ell_n)\rangle$ can be constructed, taking into account synchronisations with $\Ps \setminus \Ps'$.  Whenever a worker thread can be launched, \HIVE\ selects a swarm trace. When the \HIVE\ tool is launched to start an ISV, this is done first.

A launched worker thread performs an informed swarm search (\ISS), implemented in \LTSmin\ (Alg.~\ref{alg:swarmsearch} and F2).
In Alg.~\ref{alg:swarmsearch}, $\sigma$ is the swarm trace assigned by the \HIVE\ tool, and $\sigma(i)$ is
the singleton set containing the $(i+1)^{\it th}$ element of $\sigma$ (If $\sigma$ contains fewer than $i+1$ elements, we say that $\sigma(i)= \emptyset$). In the \ISS, $\sprocess$ is explored, but not
exhaustively: the potential behaviour of the subsystem $\Ps'$ is restricted to $\sigma$, which restricts exploration
of $\sprocess$. For each visited state $s$, $\Nextset$ is extended with $\Next \mid (\actions \setminus A) (s)$, i.e.\
all successor states reachable via behaviour of $\Ps \setminus \Ps'$, and $\Step$ is extended with $\Next \mid \sigma(i) (s)$,
i.e.\ all successor states reachable via the current behaviour in $\sigma$.

\begin{wrapalgorithm}[13]{l}{0.32\textwidth}
{\scriptsize
\caption{\BFS-based \textsc{Iss}}
\begin{algorithmic}\label{alg:swarmsearch}
\REQUIRE $\Ps$, $\sint$, $\actions$, $A = \actions' \cup \M(\actions')$, $\sigma$
\ENSURE $\sprocess$ restricted to $\sigma$ is explored
\STATE $i\leftarrow 0$
\STATE $\Open\leftarrow \sint$; $\Closed, \Nextset, \Step, \feedback_i \leftarrow \emptyset$
\WHILE{$\Open\neq\emptyset \vee \Step \neq\emptyset$}
\IF{$\Open = \emptyset$}
\STATE $i \leftarrow i+1$
\STATE $\Open\leftarrow \Step \setminus \Closed$; $\Step, \feedback_i \leftarrow \emptyset$
\ENDIF
\FORALL {$s \in \Open$}
\STATE $\Nextset \leftarrow \Nextset \cup \Next \mid (\actions \setminus A) (s)$
\STATE $\Step \leftarrow \Step \cup \Next \mid \sigma(i) (s)$
\STATE $\feedback_i \leftarrow \feedback_i \cup \{\ell \in A \mid \Next \mid \{\ell\} (s) \neq \emptyset \}$
\ENDFOR
\STATE $\Closed \leftarrow \Closed \cup \Open$
\STATE $\Open \leftarrow \Nextset \setminus \Closed$; $\Nextset \leftarrow \emptyset$
\ENDWHILE
\end{algorithmic}
}
\end{wrapalgorithm}
When all states in $\Open$ are explored, the contents
of $\Nextset$ is moved to $\Open$, after duplicate detection (for which the search history $\Closed$ is used). Note that when all reachable states have been explored in
this manner, $i$ is increased, by which the \ISS\ moves to the next step in $\sigma$, and new states become available.
The main idea of ISV is to construct the set of all possible traces through $\sprocess'$, and to perform an \ISS\ through $\sprocess$ for each
of those traces. This means that eventually $\sprocess$ is completely explored. A proof of correctness can be sketched as follows: say that
all traces through $\sprocess'$ have been used by workers to explore $\sprocess$, and that after this, some reachable state $s\in\states$ has never been
visited. We will show that this leads to a contradiction. It follows from Alg.~\ref{alg:swarmsearch} that for each state $t$ to be explored,
all new states $t' \in \Next \mid (\actions \setminus A) (t)$ are going to be explored as well, and for some $i$, $t'' \in \Next \mid \sigma(i) (t)$ is going to be
added to $\Step$. This implies that all states ${\hat t} \in \Next \mid (A \setminus \sigma(i)) (t)$ are going to be ignored. From this and the fact that $\sint$ is explored, it follows that a state
$s$ is ignored iff for all traces through $\sprocess$ from $\sint$ to $s$, there exist $t, {\hat t} \in \states$ such that $\sint \tstep{\actions}{^*} t \step{\ell} {\hat t} \tstep{\actions}{^*} s$, with $\ell \in A \setminus \sigma(i)$, $i$ being the current position
in $\sigma$ when exploring $t$. Let us consider one of those traces. We call $\sigma'$ the swarm trace followed to reach $t$ from $\sint$ over that trace. Note that this is a prefix of $\sigma$.
Let us assume that by following $\sigma'$ extended with $\ell$, $s$ can be reached from $\sint$.\footnote{This is not true if there are multiple transitions stemming from $\Ps'$ on the trace to $s$ not agreeing with $\sigma$, but then, we can repeat the reasoning in the proof sketch until there is only one left.} Since $\sigma'$ has been derived from a trace through $\sprocess'$ and $\ell \in A$, the extended trace must also be derivable from a trace through $\sprocess'$. But then, since all traces through $\sprocess'$ have been used in the ISV, $s$ must have been visited by some other worker that followed $\sigma'$, and we have a contradiction.

In case $\Ps$ and $\Ps'$
sometimes synchronise, the trace counting will produce an over-approximation of the possible set of traces of $\Ps'$
{\it in the context of} $\Ps$. This is because in the trace counting, it is always assumed that whenever $\Ps'$ needs to synchronise,
this can happen in $\Ps$. The result is that some swarm traces may not correspond with actual potential behaviour in $\sprocess$. To deal with
this, \ISS\ includes a feedback procedure: For every position $i$ in $\sigma$, it is recorded in $\feedback_i$ which potential behaviour
of $\Ps'$ has actually been observed. When finished, \ISS\ returns the $\feedback_i$, and using these, \HIVE\ can prune away
both $\sigma$ and other, invalid, traces (F3). Since each trace prefix represents a set of traces with {\it consecutive numbers}
(see e.g.\ the ranges for states in trace 10 in Fig.~\ref{fig:ipod}), the set of explored and pruned swarm traces can be represented in a relatively small list of ranges.
Initially, the set of swarm traces is empty. Say we explore the
\LTS\ $\sprocess$ of the DRM specification, and $\sprocess'$ is as displayed in Fig.~\ref{fig:ipod}, and say it is detected that synchronisation with
\textsf{recv\_C1\_2} at state $0$ (to state $3$) can actually not happen
in $\sprocess$. As already mentioned, $\langle \textsf{recv\_C1\_2} \rangle$ represents $[6, 13\rangle$. So after pruning, $[6,13\rangle$ represents
the new set of explored traces. Furthermore, elements in the list can often be merged. E.g., if ranges $[0,5\rangle$ and $[8,14\rangle$ have been explored earlier, and range $[5,8\rangle$ is to be added, the result can again be described using a single range $[0,14\rangle$. At all times, \HIVE\ is ready to launch another worker (F1) and to prune more traces (F3). When there are no more swarm traces
left to explore, the ISV is finished.

\section{Implementation and Using the Tools}
\label{sec:impl}

\paragraph*{Implementation}

The trace-counting \DFS\ and \ISS\ have both been implemented in an unofficial extension of the \LTSmin\ toolset version 1.6-19, which has been written in $C$. Since \LTSmin\ already contains a whole range of exploration
algorithms (both explicit-state and symbolic), there was no need to implement new data structures. ISV is very light-weight in terms of communication between the \HIVE\ and the workers,
the only information sent to launch an \ISS\ being a swarm trace in the form of a list of actions. This list is being stored in \LTSmin\ in a linked list, and a pointer traverses this list when exploring,
to keep track of the current swarm trace position. In addition to this, a bit set is used to keep track of the encountered actions stemming from $\Ps'$ since the last move along the swarm trace.
This is done to construct the $\feedback_i$. A bit set implementation using a tree data structure is available in the \LTSmin\ toolset.

Unfortunately, it is currently not possible to automatically extract $\Ps'$ of a given subsystem from $\Ps$, meaning that $\Ps'$ must manually be derived from $\Ps$. At times, this requires quite some inside knowledge of the description, therefore it is at the moment the main reason that we have not yet performed more experiments. Automatic construction of the $\Ps'$ description is listed as future work (see Section~\ref{sec:conc}).

The \HIVE\ tool consists of about 1,200 lines of $C$-code. Because of the communication being light-weight, and because interactions between the workers and \HIVE\ either involve asking for a new trace and receiving it, or sending the results of an \ISS, we decided to implement all communication in the request-response method using TCP/IP sockets. During an ISV, \HIVE\ frequently needs to extract traces from the $\sprocess'$, which is kept in memory together with the {\it tc}-function. Besides this, a linked list $L$ of nodes containing trace ranges (the $[i,j\rangle$ mentioned in Section~\ref{sec:ISV}) is maintained, representing the set of explored traces. Currently, when a trace is selected (F1), an ID is chosen randomly from $L$, but one can imagine other selection strategies (see Section~\ref{sec:conc}). Then, the corresponding trace is extracted from $\sprocess'$.

When launching many workers, frequent requests to \HIVE\ are to be expected. Therefore, $\HIVE$ has been implemented with \textit{pthreads}; whenever a worker sends a request, a new thread is launched in \HIVE\ to handle the request. If a new swarm trace is required, F1 is performed, and if feedback is given, F3 is performed. The \LTS\ $\sprocess'$ is never changed, hence no race conditions can occur when multiple threads read it, but $L$ is frequently accessed and updated, when selecting a trace and processing feedback, the latter involving writing. For this reason, we introduced a data lock on $L$. We plan to use more fine-grained locking in the future, but we have not yet experienced a real slowdown when using one lock.

During an ISV, \HIVE\ keeps accepting new requests until $L$ has one node containing the range $[0, {\it tc}(\sint') \rangle$. From that moment on, any requests are answered with the command to terminate, effectively ending all worker executions. The same is done if a worker reports in its feedback that a counter-example to a property to check has been found, because the ISV can stop immediately in that case.

Finally, all has been tested on \textsc{Linux} (\textsc{Red Hat} 4.3.2-7 and \textsc{Debian} 6.0.1) and \textsc{Mac OS X} 10.6.8.

\paragraph*{Setting up and launching an ISV}

In the following, we assume that we have a $\mu$CRL specification named \textsf{spec.mcrl} describing $\Ps$, and a specification named \textsf{specsub.mcrl} describing $\Ps'$. Actually, any action-based modelling language compatible with \LTSmin\ is suitable for ISV as well. A $\mu$CRL specification is usually first linearised to a \textsf{tbf} file, using the $\mu$CRL toolset~\cite{blom.fokkink.groote.mcrltoolset}, which is subsequently used as the actual input of \LTSmin. Having \textsf{spec.tbf} and \textsf{specsub.tbf}, the weighted $\sprocess'$ is saved to disk as follows, with $\langle \textsf{sub} \rangle$ being the chosen base name for the files storing the weighted $\sprocess'$:\footnote{For $\mu$CRL specifications, \textsf{lpo2lts-grey} is the explicit-state space generator of \LTSmin. For other modelling languages, another appropriate \LTSmin\ tool should be used.}

\[\textsf{lpo2lts-grey --getswarm=}\langle \textsf{sub}\rangle\ \textsf{specsub.tbf}\]

This produces the files \textsf{sub.swh}, \textsf{sub.swc}, and \textsf{sub.sww}, containing the actions in $\actions'$, the transitions in $\transitions'$ (with actions and states represented by numbers), and the weights of the states, respectively. Actually, if \textsf{specsub.tbf} yields infinite behaviour, this can be bounded by a depth $n$ using the option $\textsf{--swbound=}\langle \textsf{n}\rangle$.

The \HIVE\ can now be launched on the same machine by invoking the following, with $\langle \textsf{portnr}\rangle$ being the port number it is supposed to listen at for incoming requests:

\[\textsf{hive } \langle \textsf{portnr}\rangle\ \langle \textsf{sub}\rangle\]

An \ISS\ can be started as follows, $\langle \textsf{server} \rangle$ being the IP address of the machine running \HIVE:

\[\textsf{lpo2lts-grey --swarm=}\langle \textsf{sub}\rangle \textsf{ --hiveserver=}\langle \textsf{server}\rangle \textsf{ --hiveport=}\langle \textsf{port}\rangle \textsf{ spec.tbf}\]

Note that each \ISS\ also needs information on $\sprocess'$. Actually, only \textsf{sub.swh} is read from disk, to learn $\actions'$. Therefore, this file should be available on all machines where $\ISSs$ are started. Finally, in practice, one often wants to start many \ISSs\ simultaneously, and start a new \ISS\ every time one terminates. This whole procedure can be launched using the shell script \textsf{hive\_launch.sh}.

\section{Experimental Results}
\label{sec:exp}

Table~\ref{tab:results} shows experimental results using the $\mu$CRL~\cite{gp95} specifications of a DRM procotol~\cite{drm} and the Link Layer Protocol of the IEEE-1394 Serial Bus (Firewire)~\cite{1394link}. We were not yet able to perform more experiments using other specifications, mainly because subsystem specifications still need to be constructed manually, which requires a deep understanding of the system specifications. The experiments were performed on a machine with two dual-core \textsc{amd opteron} (tm) processors 885 2.6 GHz, 126 GB RAM, running \textsc{Red Hat} 4.3.2-7. We simulated the presence of 10 and 100 workers for each experiment (the fully independent worker threads can also be run in sequence). This has some effect on the results: in order to simulate 10 and 100 parallel \ISSs, \HIVE\ postponed the processing of \ISS\ feedback until 10 and 100 of them had been accumulated, respectively. When the \ISSs\ truly run in parallel, this feedback processing is done continuously, and redundant work can therefore be avoided at an earlier stage. In the DRM case, we selected both one and two iPod processes for $\Ps'$, and in the Firewire case, a bounded analysis of one of the link protocol entities resulted in $\sprocess'$. The SV runs have been performed with the \DFS\ of \LTSmin. Since the specifications are correct, there is no early termination for the explorations, meaning that in SV, all reachable states are explored 10 times. In the DRM case, ISV based on one iPod process leads to an initial swarm set with 5,124 traces, 45 of which were actually needed for different runs. Each run needed to explore {\it at most} 18\% of $\sprocess$, and in total, the number of states explored was smaller than in the SV. ISV based on two iPod processes leads to a much larger swarm set, and clearly, feedback information is essential. ISV with 10 parallel workers explored in total $2.5$ times more states than SV, but each \ISS\ covered at most $\frac{1}{2}\%$ of $\sprocess$, meaning that they needed a small amount of memory. This demonstrates that ISV is useful in a network where the machines do not have large amounts of RAM. In the Firewire case with 10 parallel workers, each \ISS\ explored at most only $\frac{1}{6}\%$ of $\sprocess$, and in total, the SV explored $83\%$ more states than the ISV. In terms of scalability related to the number of parallel workers, the results with 100 workers show that the overall execution times can be drastically reduced when increasing the number of workers: compared to having 10 workers, 100 workers reduce the time by $86\%$ in the DRM case, and $88\%$ in the Firewire case. The number of \ISSs\ has actually increased, but we expect this to be an effect of the simulations of parallel workers, as explained before.

A full experimental analysis of the algorithms would also incorporate cases with bugs, to test the speed of detection. This is future work, but since ISV has practically no overhead compared to SV, and the \ISSs\ are embarrassingly parallel and explore very different parts of a state space, we expect ISV and SV to be comparable in their bug-hunting capabilities.
Finally, we chose not to compare ISV experimentally with other distributed techniques (e.g.\ those using frequent synchronisations), because there are too many undesired factors playing a role when doing that (e.g.\ implementation language, modelling language, level of expertise of the user with the model checker).

\begin{table}[t]
\begin{small}
\begin{center}
\caption{Results for bug-free cases with SV and ISV, 10 and 100 workers.}
\label{tab:results}
\begin{tabular}[c]{|clrrrrrrr|} \hline
{\tabsize {\bf case}} & {\tabsize {\bf \# workers}} & {\tabsize {\bf search}} & \multicolumn{6}{c|}{\tabsize {\bf results}} \\
{\tabsize } &  & & {\tabsize {\it \# est.\ runs}} & {\tabsize {\it \# runs}} & {\tabsize {\it max.\ \# states}} & {\tabsize {\it max.\ time}} & {\tabsize {\it total \# states}} & {\tabsize {\it total time}}\\
\hline\hline
\multirow{4}{*}{{\tabsize DRM (1nnc, 3ttp)}} & 10 & {\tabsize SV} & {\tabsize 10} & {\tabsize 10} & {\tabsize 13,246,976} & {\tabsize 19,477 s} & {\tabsize 132,469,760} & {\tabsize 19,477 s}\\
& 10 & {\tabsize ISV, 1 iPod} & {\tabsize $5,124$} & {\tabsize 45} & {\tabsize 2,352,315} & {\tabsize 2,832 s} & {\tabsize 85,966,540} & {\tabsize 14,157 s}\\
& 10 & {\tabsize ISV, 2 iPods} & {\tabsize $1.31*10^{13}$} & {\tabsize 7,070} & {\tabsize 70,211} & {\tabsize 177 s} & {\tabsize 353,591,910} & {\tabsize 125,139 s}\\
& 100 & {\tabsize ISV, 2 iPods} & {\tabsize $1.31*10^{13}$} & {\tabsize 9,900} & {\tabsize 70,211} & {\tabsize 175 s} & {\tabsize 361,050,900} & {\tabsize 17,325 s}\\
\hline\hline
\multirow{3}{*}{{\tabsize 1394 (3 link ent.)}} & 10 & {\tabsize SV} & {\tabsize 10} & {\tabsize 10} & {\tabsize 137,935,402} & {\tabsize 105,020 s} & {\tabsize 1,379,354,020} & {\tabsize 105,020 s}\\
& 10 & {\tabsize ISV} & {\tabsize $3.01*10^9$} & {\tabsize 1,160} & {\tabsize 236,823} & {\tabsize 524 s} & {\tabsize 235,114,520} & {\tabsize 60,784 s}\\
& 100 & {\tabsize ISV} & {\tabsize $3.01*10^9$} & {\tabsize 1,400} & {\tabsize 236,823} & {\tabsize 521 s} & {\tabsize 252,206,430} & {\tabsize 7,294 s}\\
\hline
\end{tabular}
\linebreak[4]
\linebreak[4]
{\footnotesize  \# est.\ runs: estimated \# runs needed (\# of swarm traces for ISV). \# runs: actual \# runs needed. {\it total (max.) \# states}: total (largest) \# states explored (in a single search). {\it total (max.) time}: total (longest) running time (of a single search).}
\end{center}
\end{small}
\end{table}

\section{Conclusions and Future Work}
\label{sec:conc}

We presented the functionality of the \HIVE\ tool and two new \LTSmin\ algorithms for ISVs. ISV is an SV method for action-based formalisms to bound the embarrassingly parallel workers to different \LTS\ parts. Worst case, if the system under verification is correct, no worker needs to perform an exhaustive exploration, and memory and time requirements for a single worker can remain low.

{\bf Tool availability} Both the ISV extended version of \LTSmin\ and \HIVE\ are available at \url{http://www.win.tue.nl/~awijs/suppls/hive_ltsmin.html}.

{\bf Future work} We plan to further develop the \HIVE\ tool such that a description $\Ps'$ of a given subsystem can automatically be derived from a given description $\Ps$. We also wish to investigate which kind of subsystems are particulary effective for the work distribution in ISV, and which are not, so that an automatic subsystem selection method can be derived. If $\Ps'$ yields infinite behaviour, we want to support its full behaviour automatically in the future. As long as $\sprocess$ is finite-state, this should be possible. Furthermore, we want to investigate different strategies to select swarm traces and to guide individual \ISSs. Finally, we will perform more experiments with much larger state spaces, using a computer cluster.

\bibliographystyle{eptcs}
\bibliography{shortlit}
\fi
\end{document}

%% file: preamble.tex
\newcommand{\bisimilar}[1][]{%
  \setbox0=\hbox{\kern-.1ex{$\leftrightarrow$}\kern-.1ex}
  \setbox1=\vbox{\hbox{\raise .1ex \box0}\hrule}%
  \ensuremath{\mathrel{\hbox{\kern.1ex\box1\kern.1ex}_{#1}}}
}

\newcommand{\natdom} {\ensuremath{\mathds{N}}}

\renewcommand{\min}{\mathit{min}}

\newcommand{\sprocess} {\ensuremath{{\mathcal{P}}}}
\newcommand{\states} {\ensuremath{{\mathcal{S}}}}

\newcommand{\transitions} {\ensuremath{{\mathcal{T}}}}
\newcommand{\actions} {\ensuremath{{\mathcal{A}}}}

\newcommand{\Open} {\ensuremath{\mathit{Open}}}
\newcommand{\Closed} {\ensuremath{\mathit{Closed}}}
\newcommand{\Nextset}{\ensuremath{\mathit{Next}}}
\newcommand{\Step}{\ensuremath{\mathit{Step}}}

\newcommand{\feedback} {\ensuremath{\mathit{F}}}

\newcommand{\Next}{\mathcal{N}}

\newcommand{\tstep}[2]{\ensuremath{\mathbin{\stackrel{#1}{\longrightarrow}_{#2}}}}
\newcommand{\step}[1]{\ensuremath{\mathbin{\stackrel{#1}{\longrightarrow}}}}





%% file: hivetool.bbl
\begin{thebibliography}{10}
\providecommand{\bibitemdeclare}[2]{}
\providecommand{\urlprefix}{Available at }
\providecommand{\url}[1]{\texttt{#1}}
\providecommand{\href}[2]{\texttt{#2}}
\providecommand{\urlalt}[2]{\href{#1}{#2}}
\providecommand{\doi}[1]{doi:\urlalt{http://dx.doi.org/#1}{#1}}
\providecommand{\bibinfo}[2]{#2}

\bibitemdeclare{inproceedings}{divine10}
\bibitem{divine10}
\bibinfo{author}{J.~Barnat}, \bibinfo{author}{L.~Brim},
  \bibinfo{author}{M.~Ce\v{s}ka} \& \bibinfo{author}{P.~Ro\v{c}kai}
  (\bibinfo{year}{2010}): \emph{\bibinfo{title}{Di{V}in{E}: {Parallel}
  {D}istributed {M}odel {C}hecker}}.
\newblock In: {\sl \bibinfo{booktitle}{HiBi/PDMC'10}}, pp.
  \bibinfo{pages}{4--7}, \doi{10.1109/PDMC-HiBi.2010.9}.

\bibitemdeclare{inproceedings}{preach}
\bibitem{preach}
\bibinfo{author}{B.~Bingham}, \bibinfo{author}{J.~Bingham},
  \bibinfo{author}{F.M. de~Paula}, \bibinfo{author}{J.~Erickson},
  \bibinfo{author}{G.~Singh} \& \bibinfo{author}{M.~Reitblatt}
  (\bibinfo{year}{2010}): \emph{\bibinfo{title}{Industrial {S}trength
  {D}istributed {E}xplicit {S}tate {M}odel {C}hecking}}.
\newblock In: {\sl \bibinfo{booktitle}{HiBi / PDMC 2010}},
  \bibinfo{publisher}{IEEE}, pp. \bibinfo{pages}{28--36},
  \doi{10.1109/PDMC-HiBi.2010.13}.

\bibitemdeclare{inproceedings}{blom.fokkink.groote.mcrltoolset}
\bibitem{blom.fokkink.groote.mcrltoolset}
\bibinfo{author}{S.C.C. Blom}, \bibinfo{author}{W.J. Fokkink},
  \bibinfo{author}{J.F. Groote}, \bibinfo{author}{I.~{van} Langevelde},
  \bibinfo{author}{B.~Lisser} \& \bibinfo{author}{J.C. {van}~{de} Pol}
  (\bibinfo{year}{2001}): \emph{\bibinfo{title}{{$\mu$CRL}: {A} {T}oolset for
  {A}nalysing {A}lgebraic {S}pecifications}}.
\newblock In: {\sl \bibinfo{booktitle}{CAV'01}}, {\sl \bibinfo{series}{LNCS}}
  \bibinfo{volume}{2102}, \bibinfo{publisher}{Springer}, pp.
  \bibinfo{pages}{250--254}, \doi{10.1007/3-540-44585-4_23}.

\bibitemdeclare{inproceedings}{ltsmintool}
\bibitem{ltsmintool}
\bibinfo{author}{S.C.C. Blom}, \bibinfo{author}{J.C.~{van de} Pol} \&
  \bibinfo{author}{M.~Weber} (\bibinfo{year}{2010}):
  \emph{\bibinfo{title}{$\LTSmin$: {D}istributed and {S}ymbolic
  {R}eachability}}.
\newblock In: {\sl \bibinfo{booktitle}{CAV'10}}, {\sl \bibinfo{series}{LNCS}}
  \bibinfo{volume}{6174}, pp. \bibinfo{pages}{354--359},
  \doi{10.1007/978-3-642-14295-6_31}.

\bibitemdeclare{inproceedings}{dwyer.elbaum.parallelrandomized}
\bibitem{dwyer.elbaum.parallelrandomized}
\bibinfo{author}{M.B. Dwyer}, \bibinfo{author}{S.G. Elbaum},
  \bibinfo{author}{S.~Person} \& \bibinfo{author}{R.~Purandare}
  (\bibinfo{year}{2007}): \emph{\bibinfo{title}{Parallel {R}andomized
  {S}tate-space {S}earch}}.
\newblock In: {\sl \bibinfo{booktitle}{ICSE'07}}, \bibinfo{publisher}{IEEE},
  pp. \bibinfo{pages}{3--12}, \doi{10.1109/ICSE.2007.62}.

\bibitemdeclare{book}{foster.parallel}
\bibitem{foster.parallel}
\bibinfo{author}{I.~Foster} (\bibinfo{year}{1995}):
  \emph{\bibinfo{title}{Designing and {B}uilding {P}arallel {P}rograms}}.
\newblock \bibinfo{publisher}{Addison-Wesley}.

\bibitemdeclare{inproceedings}{Garavel2006}
\bibitem{Garavel2006}
\bibinfo{author}{H.~Garavel}, \bibinfo{author}{R.~Mateescu},
  \bibinfo{author}{D.~Bergamini}, \bibinfo{author}{A.~Curic},
  \bibinfo{author}{N.~Descoubes}, \bibinfo{author}{C.~Joubert},
  \bibinfo{author}{I.~Smarandache-Sturm} \& \bibinfo{author}{G.~Stragier}
  (\bibinfo{year}{2006}): \emph{\bibinfo{title}{{DISTRIBUTOR} and {BCG\_MERGE}:
  {T}ools for {D}istributed {E}xplicit {S}tate {S}pace {G}eneration}}.
\newblock In: {\sl \bibinfo{booktitle}{TACAS'06}}, {\sl \bibinfo{series}{LNCS}}
  \bibinfo{volume}{3920}, \bibinfo{publisher}{Springer}, pp.
  \bibinfo{pages}{445--449}, \doi{10.1007/11691372_30}.

\bibitemdeclare{inproceedings}{gp95}
\bibitem{gp95}
\bibinfo{author}{J.F. Groote} \& \bibinfo{author}{A.~Ponse}
  (\bibinfo{year}{1995}): \emph{\bibinfo{title}{The {S}yntax and {S}emantics of
  $\mu${CRL}}}.
\newblock In: {\sl \bibinfo{booktitle}{ACP'94}}, \bibinfo{publisher}{Springer},
  pp. \bibinfo{pages}{26--62}.

\bibitemdeclare{inproceedings}{holzmann.swarm2}
\bibitem{holzmann.swarm2}
\bibinfo{author}{G.J. Holzmann}, \bibinfo{author}{R.~Joshi} \&
  \bibinfo{author}{A.~Groce} (\bibinfo{year}{2008}):
  \emph{\bibinfo{title}{Swarm {V}erification}}.
\newblock In: {\sl \bibinfo{booktitle}{ASE'08}}, \bibinfo{publisher}{IEEE}, pp.
  \bibinfo{pages}{1--6}, \doi{10.1109/ASE.2008.9}.

\bibitemdeclare{inproceedings}{lerda.sista.distributedspin}
\bibitem{lerda.sista.distributedspin}
\bibinfo{author}{F.~Lerda} \& \bibinfo{author}{R.~Sista}
  (\bibinfo{year}{1999}): \emph{\bibinfo{title}{Distributed-{M}emory Model
  Checking with {SPIN}}}.
\newblock In: {\sl \bibinfo{booktitle}{{SPIN}'99}}, {\sl
  \bibinfo{series}{LNCS}} \bibinfo{volume}{1680},
  \bibinfo{publisher}{Springer}, pp. \bibinfo{pages}{22--39},
  \doi{10.1007/3-540-48234-2_3}.

\bibitemdeclare{techreport}{1394link}
\bibitem{1394link}
\bibinfo{author}{S.P. Luttik} (\bibinfo{year}{1997}):
  \emph{\bibinfo{title}{Description and {F}ormal {S}pecification of the {L}ink
  {L}ayer of {P}1394}}.
\newblock \bibinfo{type}{SEN-R} \bibinfo{number}{9706},
  \bibinfo{institution}{CWI}.

\bibitemdeclare{article}{moore}
\bibitem{moore}
\bibinfo{author}{G.E. Moore} (\bibinfo{year}{1998}):
  \emph{\bibinfo{title}{{Cramming more Components onto Integrated Circuits}}}.
\newblock {\sl \bibinfo{journal}{Proc.\ of the IEEE}}
  \bibinfo{volume}{86}(\bibinfo{number}{1}), pp. \bibinfo{pages}{82--85},
  \doi{10.1109/JPROC.1998.658762}.

\bibitemdeclare{article}{drm}
\bibitem{drm}
\bibinfo{author}{M.~{Torabi Dashti}}, \bibinfo{author}{S.~Krishnan Nair} \&
  \bibinfo{author}{H.L. Jonker} (\bibinfo{year}{2008}):
  \emph{\bibinfo{title}{Nuovo {DRM} {P}aradiso: {T}owards a {V}erified {F}air
  {DRM} {S}cheme}}.
\newblock {\sl \bibinfo{journal}{Fundamenta Informaticae}}
  \bibinfo{volume}{89}(\bibinfo{number}{4}), pp. \bibinfo{pages}{393--417}.

\bibitemdeclare{inproceedings}{wijs.isv}
\bibitem{wijs.isv}
\bibinfo{author}{A.J. Wijs} (\bibinfo{year}{2011}):
  \emph{\bibinfo{title}{Towards {I}nformed {S}warm {V}erification}}.
\newblock In: {\sl \bibinfo{booktitle}{NFM'11}}, {\sl \bibinfo{series}{LNCS}}
  \bibinfo{volume}{6617}, \bibinfo{publisher}{Springer}, pp.
  \bibinfo{pages}{422--437}, \doi{10.1007/978-3-642-20398-5_30}.

\end{thebibliography}
